\documentclass[12pt]{article}
\usepackage{amssymb}
\usepackage{epsfig}
\usepackage{graphicx}
\usepackage{amsmath}
\usepackage{verbatim}

\csname @addtoreset\endcsname{equation}{section}

\begin{document}
\begin{titlepage}
\title{\bf\Large Constrained Superfields and Standard Realization of Nonlinear Supersymmetry  \vspace{18pt}}

\author{\normalsize Hui~Luo,~Mingxing~Luo and Sibo~Zheng  \vspace{12pt}\\
{\it\small Zhejiang Institute of Modern Physics, Department of Physics,}\\
{\it\small Zhejiang University, Hangzhou 310027, P.R. China}\\
}

\date{}
\maketitle \voffset -.3in \vskip 1.cm \centerline{\bf Abstract}
\vskip .3cm
A constrained superfield formalism has been proposed in \cite{Seiberg} to analyze the low energy physics related to Goldstinos.
We prove that this formalism can be reformulated in the language of standard realization of nonlinear supersymmetry.
New relations have been uncovered in the standard realization of nonlinear supersymmetry.

\vskip 5.cm \noindent October  2009
 \thispagestyle{empty}

\end{titlepage}
\newpage

Associated with spontaneous breaking of global symmetries, there are always massless Nambu-Goldstone (NG) particles.
The properties of these NG particles depend on the nature of the broken and unbroken symmetries.
In strong interactions, pions are closely related to the spontaneous breaking of chiral symmetry.
The low energy physics related to the pions has been systematically analyzed in the framework of nonlinear realization of chiral symmetry
(see \cite{Weinberg} and the reference therein).

For spontaneous breaking of global supersymmetry (SUSY), one gets a fermionic NG particle, the Goldstino.\footnote{
In the presence of supergravity, the Goldstino becomes part of the massive gravitino.
However, if SUSY breaks at a scale much smaller than the Planck scale, the lower energy physics will be dominated by the Goldstino.
Depending on the detailed nature of SUSY breaking, Goldstino physics could be of importance at the TeV scale, as to be tested in the coming LHC experiments.}
The low energy physics related to the Goldstino has also been studied in the framework of nonlinear realization of SUSY.
Such a framework was first proposed by \cite{VA} and most of the study was formalized in the so-called standard realization of nonlinear SUSY \cite{IK78,IK77,IK82}.
Both superfield \cite{Wess83} and component formalisms \cite{CL04,CL96} have been developed.\footnote{
To reconcile gauge symmetries with SUSY,
gauge fields have been assigned a slightly different transformation rule than that in the standard realization \cite{Klein}.
There will be more discussions on this issue at the end of the paper.}

Recently, a new approach has been proposed to address the low energy physics related to Goldstinos \cite{Seiberg}.
Instead of a manifestly nonlinear realization of SUSY, constrained superfields are used to integrate out heavy components.
The Goldstino resides in a (constrained) chiral superfield $X_{NL}$.
The standard superspace technique is retained to write out Lagrangians while the superfields are constrained to include only the light degrees of freedom.
In this paper, we will prove that such a procedure can be reformulated in the language of standard realization of nonlinear SUSY.
The chiral superfield $X_{NL}$ can be constructed from the Goldstino and a suitable matter field.
The constraints on other superfields in \cite{Seiberg} can also be rephrased in the language of the standard realization.

In the standard realization of nonlinear SUSY, the Goldstino field and matter fields change as,
\begin{equation}{\label{standard1}}
\delta_\xi \tilde{\lambda}_{\alpha}
=\frac{\xi_{\alpha}}{\kappa}- i \kappa v^\mu_\xi \partial_{\mu}\tilde{\lambda}_{\alpha}, \ \ \
\delta_\xi \bar{\tilde{\lambda}}_{\dot\alpha}
=\frac{\bar{\xi}_{\dot\alpha}}{\kappa}- i \kappa v^\mu_\xi \partial_{\mu} \bar{\tilde{\lambda}}_{\dot\alpha}
\end{equation}
\begin{equation}
\delta_\xi \tilde \varphi = -i\kappa v^\mu_\xi \partial_{\mu} \tilde\varphi
{\label{standard2}}
\end{equation}
under a SUSY transformation, respectively.
Here $v^\mu_\xi = \tilde{\lambda}\sigma^\mu \bar{\xi} -\xi\sigma^\mu\bar{\tilde{\lambda}}$.
For discussions related to chiral fields, as in most of this paper,
it is more convenient to use the chiral version of nonlinear SUSY.
The transformations are defined as,
\begin{equation}{\label{chiral1}}
\delta_\xi \lambda_\alpha
={\xi_\alpha \over \kappa}- 2i\kappa\lambda\sigma^\mu \bar\xi \partial_\mu \lambda_\alpha, \ \ \
\delta_\xi \bar\lambda_{\dot\alpha}
={\bar\xi_{\dot\alpha} \over \kappa} + 2 i \kappa \xi \sigma^\mu \bar\lambda \partial_{\mu} \bar{\lambda}_{\dot\alpha}
\end{equation}
\begin{equation}{\label{chiral2}}
\delta_{\xi}\varphi=-2i\kappa\lambda\sigma^\mu \bar{\xi}\partial_\mu \varphi, \ \ \
\delta_{\xi}\varphi^c=2i\kappa \xi \sigma^\mu \bar{\lambda}\partial_\mu \varphi^c
\end{equation}
They are equivalent and related to the non-chiral one, via,
\begin{eqnarray}
\lambda_{\alpha}(x)=\tilde{\lambda}_{\alpha}(z), &
\bar\lambda_{\dot\alpha}(x)=\bar{\tilde{\lambda}}_{\dot\alpha}(z^*) \nonumber \\
z =x-i\kappa^{2}\tilde{\lambda}(z)\sigma \bar{\tilde{\lambda}}(z), &
z^* =x +i\kappa^{2}\tilde{\lambda}(z^*)\sigma \bar{\tilde{\lambda}}(z^*) {\label{T}} \\
\varphi(x) = \tilde\varphi(z), & \varphi^c(x) = \tilde\varphi(z^*) \nonumber
\end{eqnarray}

It started in \cite{Seiberg}  with the assumption that the Goldstino field resides in a chiral superfield $X_{NL}$,
following the supersymmetry structure and its breaking.
It transforms linearly under SUSY transformations:
\begin{eqnarray}
 \delta_{\xi} x_{NL} & = & \sqrt 2 \xi G,  \nonumber \\
%\end{equation}\begin{equation}
 \delta_{\xi} G_\alpha &=& \sqrt{2} \xi_{\alpha} F + i \sqrt 2 (\sigma^\mu \bar \xi)_\alpha\partial_\mu x_{NL}~, \label{lsusy} \\
%\end{equation}\begin{equation}
 \delta_{\xi} F &=& i \sqrt 2 \bar \xi \bar \sigma^\mu \partial_\mu G_\alpha . \nonumber
\end{eqnarray}
To rid of the scalar component, an operator identity $X_{NL}^2=0$ was proposed.
This fixes $x_{NL}= G^2 / 2 F$ up to an uninteresting additive constant, and one concludes that
\begin{equation}
X_{NL}={G^2\over 2F}+\sqrt2\theta G+\theta^2 F ~, \label{xnl}
\end{equation}
where all fields are functions of $y = x + i \theta \sigma \bar\theta$.
One can easly verify that (\ref{xnl}) is consistent with the linear SUSY algebra of (\ref{lsusy}).

Now we define $\lambda^{NL} =  G/\sqrt 2 \kappa F$.
It is straightforward to verify that, according to the linear SUSY algebra of (\ref{lsusy}),
\begin{eqnarray}
\delta_{\xi}\lambda^{NL}_{\alpha}
=\frac{\xi_{\alpha}}{\kappa}-2i\kappa \lambda^{NL}\sigma^\mu \bar{\xi}\partial_\mu \lambda^{NL}_{\alpha}
\end{eqnarray}
under SUSY transformations.
That is, $\lambda^{NL}$ transforms in exactly the same way as the Goldstino $\lambda$ in the chiral version of standard realization under supersymmetry transformations.
Naturally, $\lambda^{NL}$ can be identified as the nonlinear Goldstino field and $X_{NL}$ can be written as
\begin{equation}
X_{NL} = F \Theta^2
\end{equation}
where $\Theta = \theta + \kappa \lambda$ is introduced for later convenience.

Actually, the above procedure can be reversed.
For an arbitrary chiral superfield $\Phi = \phi + \sqrt 2 \theta \psi + \theta^2 F$,
one can always define $\lambda^\Phi = \psi/\sqrt 2 \kappa F$.
From the transformation rules in (\ref{lsusy}), one easily gets
\begin{eqnarray}
\delta_\xi \lambda^\Phi_\alpha
={\xi_\alpha \over \kappa}- 2i \kappa \lambda^\Phi \sigma^\mu \bar\xi \partial_\mu \lambda^\Phi_{\alpha}
+ {i \over \kappa F}  (\sigma^\mu \bar{\xi})_\alpha  \partial_\mu \left( \phi - {\psi^2 \over 2 F} \right)
\end{eqnarray}
Demanding $\lambda^\Phi$ to transform in the same way as that of $\lambda$, one gets $\phi = \psi^2 / 2 F$ up to an additive constant.
In such case, one again obtains $\Phi^2 = 0$ as well as $\Phi = F \Theta^2$.

In \cite{Wess83}, $\lambda$ was prompted to a linear superfield by using SUSY transformations in (\ref{chiral1}),
\[ %\begin{equation}
\Lambda(\lambda)=\exp(\theta Q+\bar{\theta} \bar{Q})\times\lambda
\] %\end{equation}
Out of $\Lambda$ and its conjugate $\bar \Lambda$, one can construct two possible chiral fields
\[ %\begin{eqnarray}
\Phi_2 = -{1\over 4} \bar D^2 \Lambda \Lambda, \ \ \ \Phi_4 = -{1\over 4}  \bar D^2 \Lambda \Lambda \bar\Lambda \bar\Lambda.
\] %\end{eqnarray}
Direct calculations yield,
\begin{equation}
\Phi_2 = f_2(\lambda) \Theta^2,  \  \ \
\Phi_4 = f_4(\lambda) \Theta^2,
\end{equation}
where $f_2$ and $f_4$ are two definite functions of $\lambda$ and their explicit forms do not concerns us here.
$f_4$ is actually the Akulov-Volkov Lagrangian for the Goldstino \cite{VA} up to an overall constant and possible total derivative terms.
It is straightforward to verify that $f_4/f_2$ and $F/f_2$ transform as matter fields via (\ref{chiral2}),
by using the SUSY algebra of (\ref{lsusy}).
Such relations have not been expected.

It has been well known that $\Phi_4$ satisfies the following constraint \cite{Wess83,Rocek}
\[
\Phi_4 \bar D^2 \bar \Phi_4 \sim \Phi_4
\]
while $\Phi_2$ does not.
For long, this constraint was used as the rationale to choose $\Phi_4$ instead of $\Phi_2$ to be the superfield for Goldstino.
As one sees now, they differ only by a matter field in the standard realization.
In retrospect, this relation is rather easy to understand. Both $\Phi$'s satisfy the same algebraic constraint of $X_{NL}$,
that is, $\Phi_2^2 = \Phi_4^2= 0$.
They must have the same form of factorization.

To recapture, the constrained $X_{NL}$ in \cite{Seiberg} can be factorized as follows
\begin{equation}
X_{NL}  = F^{'} \Phi_4 = F^{'} f_4(\lambda) \Theta^2
\end{equation}
where $F^{'}$ transforms as a nonlinearly realized matter field and $\Phi_4$ is constructed from the nonlinearly realized Goldstino field alone.
Obviously, we can reverse the procedure and construct the desired $X_{NL}$ from Goldstino and an appropriate matter field.

One can also use a real superfield to describe the Goldstino \cite{Wess83}, $V_4 = \Lambda^2 \bar \Lambda^2$.
It satisfies the same constraint of $X_{NL}$, $V_4^2 = 0$. A straightforward calculation reveals a similar simple structure,
$ V_4 = \tilde f_4(\tilde \lambda) \tilde \Theta^2 \bar{\tilde \Theta}^2$, where $\tilde \Theta = \theta + \kappa \tilde \lambda$.
$\tilde f_4$ is again the Akulov-Volkov Lagrangian for the Goldstino \cite{VA} up to an overall constant and possible total derivative terms.
Note that the condition $V^2=0$ cannot be preserved under a general gauge transformation, if $V$ originates from a gauge superfield.
However, for a general real superfield $V= D \theta^2 \bar\theta^2 + \chi \theta \bar\theta^2 + \bar\chi \bar\theta \theta^2 + \cdots$,
one can also define $\lambda^V = \chi /2 \kappa D$.
From the SUSY transformation rules for a linear real superfield, one finds,
\begin{eqnarray}
\delta_\xi \lambda^V_{\alpha}
= { \xi_\alpha \over \kappa}- i ( \lambda^V\sigma^\mu \bar \xi -\xi\sigma^\mu\bar\lambda^V ) \partial_{\mu}\lambda^V_{\alpha} + {i \over D} {(\rm total \ derivatives)}
\end{eqnarray}
Demanding the total derivatives to vanish such that $\lambda^V$ transforms in the same way as that of $\tilde \lambda$, one gets $V = D \tilde \Theta^2 \bar{\tilde \Theta}^2$,
after some elementary algebra.
In this case, it is also easy to check that $D/\tilde f_4$ transforms as a matter field in the standard realization.

When SUSY is broken, some components of superfields will get heavy and need to be integrated out.
To rid of the scalar component in a chiral superfield, $Q_{NL}=\phi_q+\sqrt 2\theta \psi_q+\theta^2F_q$,
\cite{Seiberg} suggested to use the constraint,
\begin{equation}
X_{NL} Q_{NL}=0 \label{bc1}
\end{equation}
From which, one gets
\begin{equation}
\phi_q={\psi_q G \over F}-{G^2\over 2F ^2}F_q \label{bc2}
\end{equation}
To rid of the fermionic component in a chiral superfield, ${\cal H}_{NL}=  H+ \sqrt2\theta \psi_h + \theta^2 F_h$,
\cite{Seiberg} suggested to use the constraint,
\begin{equation}
X_{NL} \Bar D_{\dot\alpha} \Bar {\cal H}_{NL}=0, \ \ \
% \end{equation}
{\rm or\  equivalently,} \ \ \
%\begin{equation}
X_{NL} \Bar {\cal H}_{NL}={\rm chiral}
\end{equation}
From which, one gets,
\begin{equation}
{\cal H}_{NL}= H +  i \sqrt{2} \theta \sigma^\mu\left(\Bar G\over \Bar F\right)\partial_\mu H
+\theta^2 \left[-\partial_\nu\left({\Bar G\over \Bar F}\right)\Bar\sigma^\mu\sigma^{\nu}{\Bar G\over \Bar F}\partial_\mu H+{1\over 2\Bar F^2}\Bar G^2\partial^2 H \right]
\label{fc}
\end{equation}

In the standard realization, nonlinearly realized superfields can be obtained from linear ones via
\begin{equation}{\label{superfield}}
\hat \Phi =\exp\left[-\kappa \left(\tilde\lambda Q+\Bar{\tilde\lambda} \Bar{Q}\right)\right]\times \Phi
\end{equation}
Starting with a chiral superfield $\Phi(y,\theta)=\phi(y)+\sqrt 2\theta \psi(y)+\theta^2F_\phi(y)$ and expressing $\tilde\lambda$ in terms of $\lambda$ via (\ref{T}),
one gets
$$\hat\Phi(y,\theta,\lambda) %= \Phi(y-2i\kappa\theta \sigma \Bar\lambda(y), \theta-\kappa\lambda(y))
=\hat\phi(y,\lambda)+\sqrt 2\theta \hat\psi(y,\lambda)+\theta^2 \hat F_\phi(y,\lambda).$$
The transformation rules are as the following,
\begin{eqnarray}
\delta_{\xi}\hat \Phi=-2i\kappa\lambda\sigma^\mu \Bar{\xi}\partial_\mu \hat\Phi
\end{eqnarray}
That is, $\hat\phi$, $\hat\psi$, and $\hat F$ transform independently,
each of them can be set to vanish without in conflict with others.
Setting $\hat \phi(y,\lambda)=0$ to rid of the scalar component, one reproduces the constraint (\ref{bc2}).
Setting $\hat \psi(y,\lambda)=0$ to rid of the fermionic component, one
%\[\psi = {G\over F} F_\phi + i {\sigma^\mu\Bar G \over \bar F} \left( \partial_\mu  \phi - {G \partial_\mu\psi \over F} + {G^2\over 2 F^2} \partial_\mu F_\phi \right)\]
does not reproduce (\ref{fc}) exactly.
However, one notices by an inspection of (\ref{fc}) that
\begin{equation}
\delta_\xi H = 2 i \kappa \xi \sigma^\mu \bar\lambda \partial_\mu H
\end{equation}
which transforms in the same way of $\varphi^c$ in (\ref{chiral2}). That is, $H$ transforms as an anti-chiral matter field under SUSY transformations.
Actually, it is straightforward to verify that ${\cal H}_{NL}=\exp\left[\theta Q+\Bar\theta \Bar{Q}\right]\times H$,
by virtue of $\theta Q\times H = 2 i \kappa \theta \sigma^\mu \bar\lambda \partial_\mu H$ and $\Bar\theta \Bar{Q} \times H= 0$.

To get a real scalar from a chiral superfield, \cite{Seiberg} kept only the real component $a$ of the complex scalar as an independent degree of freedom.
The imaginary component $b$ was expressed in terms of  $a$ and $\lambda$.
This was achieved by the constraint
\begin{equation}
X_{NL} \left({\cal A}_{NL}-\bar {\cal A}_{NL} \right)=0.
\end{equation}
Its solution is of the form (\ref{fc}) with ${\cal A}_{NL}\bigr|_{\theta=0}=a+ib~,$
and $b$ is the following function of $a$ and $\lambda$
\begin{eqnarray}
b&=&{1\over 2} \left({ G\over F}\sigma^\mu{\bar G\over \bar F}\right)\partial_\mu a
-\left({i\over 8}{ G^2\over F^2}\partial_\nu\left({\bar G\over \bar F}\right)\bar\sigma^\mu\sigma^\nu{\bar G\over\bar F}\partial_\mu a+c.c.\right)\nonumber \\
&&-
{ G^2\bar G^2\over 32F^2\bar F^2}\partial_\mu\left(\bar G\over \bar F\right)\left(\bar\sigma^\rho\sigma^\mu\bar\sigma^\nu+\bar\sigma^\mu\sigma^\nu
\bar\sigma^\rho\right)\partial_\nu\left( G\over F\right)\partial_\rho a~. \label{b_sol}
\end{eqnarray}
Since ${\cal A}_{NL}$ has the form of (\ref{fc}), $a+ ib$ transforms in the same way of $\varphi^c$ in (\ref{chiral2}),
as an anti-chiral matter field under SUSY transformations.
As mentioned at the beginning of the paper, any nonlinearly realized non-chiral matter field $\tilde\varphi$
can be converted into an anti-chiral one $\varphi^c = a_\varphi^c + i b_\varphi^c$ via (\ref{T}).
Here $a_\varphi^c$ and $b_\varphi^c$ are two real fields, as functions of $\tilde\varphi$ and $\lambda$.
If $\tilde\varphi$ is a real scalar field, it is straightforward but tedious to check that $b_\varphi^c$ is related to $a_\varphi^c$ via (\ref{b_sol}).
Reversing the procedure, we get one nonlinearly realized non-chiral real matter field in the non-chiral version
from ${\cal A}_{NL}\bigr|_{\theta=0}=a+ib$ by inverting the transformation (\ref{T}).

For a supersymmetric gauge field
\begin{eqnarray}
V &= & \ c+i\theta \omega - i\bar \theta \bar \omega + \theta^2 M + \bar \theta ^2 \bar M -\theta\sigma^m \bar\theta A_m \\
&&+i \theta^2 \bar \theta (\bar \chi + {i \over 2}\bar \sigma^m \partial_m \omega) - i \bar \theta^2 \theta (\chi + {i \over 2} \sigma^m \partial_m \bar \omega)
+ {1 \over 2} \theta^2 \bar\theta^2 (D +{1 \over 2} \partial^2 c)  \nonumber
\end{eqnarray}
the nonlinearly realized version is again obtained via
\begin{equation}
\hat V =\exp\left[-\kappa \left(\tilde\lambda Q+\Bar{\tilde\lambda} \Bar{Q}\right)\right]\times V
\end{equation}
All components of $\hat V$ transform independently according to (\ref{standard2}).
In \cite{Seiberg}, a Wess-Zumino type gauge choice was obtained by imposing the constraint
$X_{NL} V_{NL}=0.$
Its components solution is
\begin{eqnarray}
c&= &{\bar G \bar\sigma^\mu G \over 2|F|^2}A_\mu   + {i\bar G^2 G\chi\over 2\sqrt2\ \bar F^2 F }
- {i G^2 \bar G \bar\chi\over  2\sqrt2F^2 \bar F } + {G^2\bar G^2D \over 8|F|^4}  + \dots \nonumber \\
\omega&=& -i{\bar G\bar \sigma^\mu \over \sqrt2\ \bar F} A_\mu + {\bar G^2\over 2\bar F^2} \chi + \dots \label{V_sol} \\
M&=& i{\ \bar G\ \bar\chi \over \sqrt2 \ \bar F} - {\bar G^2 \over 4\bar F^2} D  + \dots~, \nonumber
\end{eqnarray}
where the ellipses represent terms with more Goldstinos and more derivatives.
Interestingly, these expressions can be obtained by demanding $\hat c =  \hat \omega = \hat M = 0$,
such that $\hat V$ is in the Wess-Zumino type gauge.

To rid of the gaugino field $\chi$, it is simpler by working with the field strength superfield
\begin{eqnarray}
W_\alpha=-i\chi_\alpha+L_\alpha^\beta\theta_\beta+\sigma^m_{\alpha\dot\alpha}
\partial_m\bar\chi^{\dot\alpha} \theta^2~,\ \ \
L_\alpha^\beta = \delta^\beta_\alpha D-{i\over 2}(\sigma^m\bar\sigma^n)^\beta_\alpha F_{mn}~.
\end{eqnarray}
The constraint which eliminates the gaugino $\chi$ is $X_{NL} W_{\alpha NL}=0~.$
From this, one has \cite{Seiberg}
\begin{equation}
-i\chi_\alpha = L_\alpha^\beta {G_\beta \over \sqrt 2 F}- \sigma^m _{\alpha \dot\alpha }\partial_m  \bar \chi^{ \dot\alpha}  {G^2 \over  2F^2}
\end{equation}
This can also be obtained by imposing the condition $\hat\chi_\alpha=0$.
Here $\hat\chi_\alpha$ is the first component of $\hat W_\alpha = \exp\left[-\kappa \left(\tilde\lambda Q+\Bar{\tilde\lambda} \Bar{Q}\right)\right]\times W_\alpha$.

Finally, we comment on the compatibility between nonlinear SUSY and gauge symmetries.
When dealing with non-chiral superfields, such as the vector superfields, the chiral version of nonlinear realization loses much of its virtue.
This might be part of the reason why the constraints on vector superfields in \cite{Seiberg} are quite involved.
To deal with gauge fields, it could prove to be expedient to work with the non-chiral version.

Working in the Wess-Zumino gauge, one starts with the transformation
\begin{eqnarray}
 \delta_{\xi} A_\mu & = & -i \chi \sigma_\mu \Bar\xi + i \xi \sigma_\mu \Bar\chi \nonumber \\
%\end{equation}\begin{equation}
 \delta_{\xi} \chi_\alpha &=& \sigma^{\mu\nu} \xi_\alpha F_{\mu\nu} + i \xi D, \label{lsusy2} \\
%\end{equation}\begin{equation}
 \delta_{\xi} D &=& - D_\mu \chi \sigma^\mu  \Bar\xi - \xi \sigma^\mu D_\mu \Bar\chi . \nonumber
\end{eqnarray}
where
$$
D_\mu = \partial_\mu - i A_\mu
\ \ \ {\rm and} \ \ \
F_{\mu\nu}= \partial_\mu A_\nu - i \partial_\nu A_\mu-i [ A_\mu, A_\nu]
$$
From these transformation laws, one can construct a set of four superfields \cite{Klein}
\begin{equation}
V_\mu = \exp(\theta Q + \Bar\theta \Bar Q) \times A_\mu
\end{equation}
Out of which, we form four nonlinearly realized superfields
\begin{equation}
\hat V_\mu = \exp\left[-\kappa (\tilde \lambda Q + \Bar {\tilde\lambda} \Bar Q) \right] \times V_\mu
= \hat A_\mu + i \theta \sigma_\mu \Bar{\hat \chi} - i \hat \chi \sigma_\mu \Bar \theta + \cdots
\end{equation}
We can rid of the feminonic componet by demainding $\hat \chi = 0 $, which can be used to obtain $\chi$ in terms of $A_\mu$ and $D$.
One may proceed to get the effective Lagrangian via the superspace formalism, with such constraints.

On the other hand, it is straightforward to verify that
\begin{equation}
\delta_\xi \hat A_\mu = - i \kappa v^\nu_\xi \hat F_{\nu\mu}
\end{equation}
where $\hat F_{\mu\nu}= \partial_\mu \hat A_\nu - i \partial_\nu \hat A_\mu-i [\hat A_\mu, \hat A_\nu]$.
This can be rewritten as
\begin{equation}
\delta_\xi \hat A_\mu = - i \kappa v^\nu_\xi \partial_\nu \hat A_\mu - i \kappa \partial_\mu v^\nu_\xi \hat A_\nu + D_\mu (i\kappa v^\nu_\xi \hat A_\nu)
\end{equation}
The last term in this expression can be compensated by a gauge transformation of the parameter $-i\kappa v^\nu_\xi \hat A_\nu$.
Under this combination of SUSY and gauge transformations, one has
\begin{equation}
\delta_\xi^{'} \hat A_\mu = - i \kappa v^\nu_\xi \partial_\nu \hat A_\mu - i \kappa \partial_\mu v^\nu_\xi \hat A_\nu
\end{equation}
One may use $T_\mu^\nu = \delta_\mu^\nu - i \kappa^2 \partial_\mu \tilde\lambda \sigma^\nu \Bar{\tilde \lambda} +
i \kappa^2 \tilde\lambda \sigma^\nu \partial_\mu \Bar{\tilde\lambda}$ to define
\begin{equation}
{\cal D}_\mu =(T^{-1})^{\nu}_{\mu} D_\nu = (T^{-1})^{\nu}_{\mu} (\partial_\nu - i A_\nu)
\end{equation}
\begin{eqnarray}{\label{FV}}
{\cal F}_{\mu\nu}=(T^{-1})^\rho_\mu (T^{-1})^{\sigma}_{\nu}
(\partial_\rho A_\sigma -\partial_\sigma A_\rho - i[A_\rho ,A_\sigma])
\end{eqnarray}
Both ${\cal D}_\mu$ and ${\cal F}_{\mu\nu}$ transform covariantly under both SUSY and gauge rotation.
We can construct SUSY Lagrangians with gauge invariance in the component formalism, by simply making the substitution,
$D_\mu \rightarrow {\cal D}_\mu$ and $F_{\mu\nu} \rightarrow {\cal F}_{\mu\nu}$, in ordinary non-SUSY Lagrangians.

\section*{Acknowledgement}
This work is supported in part by the National Science Foundation
of China (10425525) and (10875103).


\begin{thebibliography} {99}

\bibitem{Weinberg}
S.~Weinberg,``The Quantum Theory of Fields,'' Vol II, Chapter 19,
Cambridge. Pr. (1996)

\bibitem{VA}
D.~V.~Volkov and V.~P.~Akulov, ``Is the Neutrino a Goldstone
Particle?,''  Phys.\ Lett.\  B {\bf 46}, 109 (1973).

\bibitem{IK78}
  E.~A.~Ivanov and A.~A.~Kapustnikov,
  ``General Relationship Between Linear And Nonlinear Realizations Of
  Supersymmetry,''
  J. Phys. A  {\bf 11}, 2375 (1978).

 \bibitem{IK77}
  E.~A.~Ivanov and A.~A.~Kapustnikov,
  ``Relation Between Linear And Nonlinear Realizations Of Supersymmetry,''
  JINR-E2-10765, Jun 1977.

\bibitem{IK82}
  E.~A.~Ivanov and A.~A.~Kapustnikov,
  ``The Nonlinear Realization Structure Of Models With Spontaneously Broken
  Supersymmetry,''
  J. Phys. G {\bf 8}, 167 (1982).

\bibitem{Wess83}
 S.~Samuel and J.~Wess,
  ``A Superfield Formulation Of The Nonlinear Realization Of Supersymmetry And
  Its Coupling To Supergravity,''
  Nucl. Phys. B {\bf 221}, 153 (1983).

\bibitem{CL04}
 T.~E.~Clark and S.~T.~Love,
  ``Nonlinear realization of supersymmetry and superconformal symmetry,''
  Phys. Rev.  D {\bf 70}, 105011 (2004),
  [arXiv:hep-th/0404162].

\bibitem{CL96}
T.~E.~Clark and S.~T.~Love,
  ``Goldstino couplings to matter,''
  Phys.\ Rev.\  D {\bf 54}, 5723 (1996)
  [arXiv:hep-ph/9608243].

\bibitem{Klein}
M. Klein, ``Couplings in pseudosupersymmetry,'' Phys.~Rev.~D~{\bf
66}, 055009~(2002), [arXiv: hep-th/0205300].

\bibitem{Seiberg}
Z. Komargodski, N. Seiberg, ``From Linear SUSY to Constrained
Superfields,'' arXiv:0907.2441.

\bibitem{Rocek}
  M.~Rocek,
  ``Linearizing The Volkov-Akulov Model,''
  Phys. Rev. Lett.  {\bf 41}, 451 (1978).



\end{thebibliography}
\end{document}